\documentclass[12pt]{iopart}
\usepackage{graphicx}
\usepackage{psfig}
\begin{document}

\def\dir{figures/}
\def\ra{\rightarrow}

\def\<{\langle}
\def\>{\rangle}
\def\({\left(}
\def\){\right)}
\def\[{\left[}
\def\]{\right]}
\def\Re{\mbox{Re}}
\def\Im{\mbox{Im}}
\def\tr{\mbox{tr}}
\def\Sgn{\mbox{sgn}}
\def\i{{\rm i}}
\def\d{{\rm d}}
\def\e{{\rm e}}

\newcommand{\al}{\alpha}
\newcommand{\be}{\beta}
\newcommand{\de}{\delta}
\newcommand{\De}{\Delta}
\newcommand{\eps}{\epsilon}
\newcommand{\ga}{\gamma}
\newcommand{\om}{\omega}
\newcommand{\Om}{\Omega}
\newcommand{\lam}{\lambda}
\newcommand{\Lam}{\Lambda}

\newcommand{\half}{{\textstyle{1 \over 2}}}

\title[Form factor of quantum graphs]
{Form factor for a family of quantum graphs:\\ An expansion to third order} 
\author{Gregory Berkolaiko\dag, Holger Schanz\ddag\ and Robert S. Whitney\S}
\address{\dag Department of Mathematics, University of Strathclyde,
  Glasgow G1 1XH, UK}
\address{\ddag Max-Planck-Institut f\"ur Str\"omungsforschung und Institut
f{\"u}r Nichtlineare Dynamik der Universit{\"a}t G{\"o}ttingen, 
Bunsenstra{\ss}e 10, D-37073 G\"ottingen, Germany}
\address{\S Theoretical Physics, University of Oxford, 1 Keble Road, 
Oxford OX1 3NP, UK}
\date{\today}
\begin{abstract}
  For certain types of quantum graphs we show that the random matrix
  form factor can be recovered to at least third order in
  the scaled time $\tau$ from periodic-orbit theory.  We consider the
  contributions from pairs of periodic orbits represented by diagrams
  with up to two self-intersections connected by up to four arcs and
  explain why all other diagrams are expected to give higher-order
  corrections only.
  
  For a large family of graphs with ergodic classical dynamics the
  diagrams that exist in the absence of time-reversal symmetry sum to zero.  
  The mechanism for this cancellation is rather general which suggests 
  that it also applies at higher orders in the expansion.  
  This expectation is in full agreement with the fact that
  in this case the linear-$\tau$ contribution, 
  the diagonal approximation, already reproduces the random matrix
  form factor for $\tau<1$.
  
  For systems with time-reversal symmetry there are more diagrams
  which contribute at third order.  We sum these contributions for
  quantum graphs with uniformly hyperbolic dynamics, obtaining
  $+2\tau^{3}$, in agreement with random matrix theory.  As in
  the previous calculation of the leading-order correction to the
  diagonal approximation we find that the third order contribution can
  be attributed to exceptional orbits representing the intersection of
  diagram classes.
\end{abstract}
\maketitle

\section{Introduction}
The recent work of Sieber and Richter \cite{SR01,S02} has renewed the
hope that spectral correlations in systems with chaotic classical
analogue can be explained within periodic-orbit theory. The
universality of these correlations, known as the BGS conjecture
\cite{BGS84}, is supported by overwhelming numerical evidence
\cite{Haake}. On the other hand there is no satisfactory theory for
individual chaotic systems, i.~e.\ without any disorder averages.
Numerically it was found that on time scales longer than the ergodic
time of the classical analogue, the fluctuations in the energy
spectrum of a quantum system follow those of an appropriate ensemble
of random matrices.  For random matrices, the form factor $K(\tau)$,
which is the Fourier transform of the spectral two-point correlator,
is
\begin{eqnarray}
  \label{kgoe}
  K_{\rm GOE}(\tau)&=&2\tau-\tau\log(1+2\tau)\qquad(0\le \tau\le 1)
  \nonumber\\
  &=& 2\tau-2\tau^{2}+2\tau^{3}+O(\tau^{4})\,,
\end{eqnarray}
for systems with time-reversal symmetry (TR), or
\begin{eqnarray}
  \label{kgue}
  K_{\rm GUE}(\tau)&=&\tau\qquad (0\le \tau\le 1)
\end{eqnarray}
for systems without time-reversal symmetry (NTR) \cite{Mehta}.

The semiclassical limit of the form factor in a quantum chaotic system can be
written in terms of a double sum over periodic orbits (PO) using the
Gutzwiller trace-formula \cite{Gut71}.  On short times the relatively small
number of contributing periodic orbits allows explicit calculation, however
the number of POs proliferates exponentially with time, so evaluating the sum
exactly quickly becomes impossible. In any case the universality of the BGS 
conjecture
suggests that beyond the ergodic time the form factor does not depend on the
specific dynamics of the given system.  Berry \cite{Ber85} explained that this
universality arises from the combined contributions of the huge number of
ergodic POs.  He then calculated the form factor, neglecting all correlations
between POs other than exact symmetries. Within this ``diagonal
approximation'', he obtained the leading order in $\tau$ of the random matrix
theory (RMT) result.  Efforts to reproduce Eqs.~(\ref{kgoe}),~(\ref{kgue})
beyond the diagonal approximation have been limited in their success.  At
present there is no way to derive the series expansion of Eq.~(\ref{kgoe})
from the POs of any chaotic system, nor is there a good explanation of
why Eq.~(\ref{kgue}) happens to be exactly reproduced by the diagonal
approximation for $\tau \le 1$.

Currently we only know how to go beyond the diagonal approximation in a few
special cases.  In \cite{SR01,S02} it was shown, 
that for uniformly hyperbolic and
time-reversal invariant billiards on surfaces with constant negative curvature
the second-order contribution $-2\tau^{2}$ is related to correlations within
pairs of orbits differing in the orientation of one of the two loops resulting
from a self-intersection of the orbit.  We went on to derive the same result
for a large family of quantum graphs \cite{KS97,KS99} with ergodic classical
dynamics, in particular our result was {\em not} restricted to uniformly
hyperbolic dynamics \cite{BSW02}. A subsequent study \cite{BHS02} indicated 
that the
mechanism generating the contribution $-2\tau^2$ also works for systems with
antiunitary symmetries other than simple time-reversal.

Given these results it is a plausible conjecture that in analogy
to disordered systems \cite{SLA98} the terms in the power series expansion of
$K(\tau)$ can be identified with the contributions of orbit pairs
generated by more and more self-intersections. In the present paper we will
explore this idea for a particular model system: Extending our recent paper
\cite{BSW02} we will calculate the form factor up to order $\tau^{3}$ for a
particular family quantum graphs.

This article is organized as follows: In Section 2 we define our model and
explain how the form factor can be expressed as a double sum over periodic
orbits. In Section 3 this sum is rewritten in terms of diagrams, representing
all orbits with a given number and topology of self-intersections.  Diagrams
resulting in a contribution of order $\tau^3$ are considered explicitly.  In
Section 4 we show that those diagrams which do not require
time-reversal invariance cancel each other.  The summation over the additional
diagrams in graphs with time-reversal invariance is performed in Section 5,
unfortunately here our results are limited to a family of graphs with
uniformly-hyperbolic classical dynamics.  Finally in section 6 we explain how
we selected the diagrams which give $\tau^3$-contributions by establishing a
heuristic rule which predicts the order of a given diagram's contribution
without an explicit calculation.


\section{Quantum graphs and periodic-orbit theory}

We consider graphs with $N$ vertices connected by a total of $B$ directed
bonds. A bond leading from vertex $m$ to vertex $l$ is denoted by $(m,l)$.
For graphs with time-reversal invariance it is necessary that for any bond
$(m,l)$ there exists also the reversed bond $(l,m)$. We do not rule out the
possibility of loops, i.e. bonds of the form $(m,m)$.
   
The discrete quantum dynamics on a graph are defined in terms of a $B\times B$
unitary time-evolution operator $S^{(B)}$, which has matrix elements
$S^{(B)}_{m'l',lm}$ describing the transition amplitudes from the directed
bond $(m, l)$ to $(l',m')$\footnote[2]{We drop the parentheses when a bond is
used as an index of a matrix.}.  The topology of the underlying graph is
reflected in the quantum dynamics because the amplitudes are non-zero only if
the two bonds are connected at a vertex, $l=l'$.  We choose
\begin{equation}
  \label{smat}
  S^{(B)}_{m'l',lm}=\delta_{l'l}\,\sigma^{(l)}_{m'm}\e^{\i \phi_{ml}}
\end{equation}
with $\sigma^{(l)}_{m'm}$ denoting the {\em vertex-scattering matrix} at
vertex $l$.  An explicit example of such a graph will be given in
Section~\ref{sect:sum_TR3}, here we keep the discussion as general as
possible. The phases $\phi_{ml}$ are random variables distributed uniformly
in $[0,2\pi]$ and define for fixed $B$ an ensemble of matrices $S^{(B)}$ which
can be used for averaging. It is possible to interpret this ensemble as an
infinite energy average for a given quantum graph with rationally independent
bond lengths \cite{KS97}. For a unitary operator like $S^{(B)}$ the form
factor is defined at integer times $t=0,1,\dots$ by
\begin{equation}
  \label{ff}
  K^{(B)}(\tau)=B^{-1}\langle|{\rm tr}S^{t}|^{2}\rangle_{\{\phi\}},
\end{equation}
where $\tau$ is the scaled time $\tau=t/B$. See \cite{Haake} for more details
on the description of two-point correlations for unitary operators.  For
finite $B$, the form factor (\ref{ff}) should be compared to ensembles of
unitary random matrices of dimension $B$ (CUE for NTR, COE for TR). However,
we are interested here in the limit of large graphs $B\to\infty$, keeping the
scaled time $\tau$ fixed
\begin{equation}
  \label{fflimit}
  K(\tau)=\lim_{B\to\infty} K^{(B)}(\tau)\,,
\end{equation}
because this is equivalent to the semiclassical limit of chaotic systems
\cite{KS97}. It is in this limit that the form factor is expected to assume
the corresponding universal form (\ref{kgoe}) or (\ref{kgue}).

Associated with the unitary matrix $S$ is the doubly stochastic matrix $M$
with 
\begin{eqnarray}
  \label{matrixM}
  M^{(B)}_{m'l,lm} = |S^{(B)}_{m'l,lm}|^2 = |\sigma^{(l)}_{m'm}|^2 \ .
\end{eqnarray}
It defines a Markov chain on the graph which represents the classical analogue
of our quantum system \cite{KS97,BG00}.  The matrix $M$ can be considered as the
Frobenius-Perron operator of the discrete classical dynamics.  Matrix elements
of powers of this operator give the classical probability to get from
bond $(m, l)$ to bond $(k, n)$ in $t$ steps 
\begin{equation}
\label{classtransprob}
P_{(m,l) \to (k,n)}^{(t)}=\left[M^t\right]_{nk,lm}\,.
\end{equation}
Under very general conditions it can be shown that the dynamics generated by
$M$ is ergodic and mixing\footnote{
  It is plausible to assume that these conditions are satisfied
  if the underlying
  graph is connected and one excludes special cases like bipartite graphs
  \cite{Norris}.}, i.~e.\ for fixed $B$ and $t\to\infty$ all transition
probabilities become equal
\begin{equation}
  \label{eq:ergodic}
  P_{(m,l) \to (k,n)}^{(t)} \to B^{-1} \mbox{ as } t\to\infty
  \qquad \forall (m,l), (k,n)\,.
\end{equation}
However, since in Eq.~(\ref{fflimit}) the limits $B\to\infty$ and $t\to\infty$ are
connected by fixing $\tau$, Eq.~(\ref{eq:ergodic}) is not sufficient for showing
agreement between PO expansion and RMT. We need a stronger condition such as
\begin{equation}
  \label{eq:ergodic-tau}
  P_{(m,l) \to (k,n)}^{(\tau B)} \to B^{-1} \mbox{ as } B\to\infty
  \qquad \forall (m,l), (k,n)\,.
\end{equation}
This was already discussed in \cite{Tan01} in connection with the diagonal
approximation.  In fact the precise condition may in principle depend on the
order to which agreement with RMT is required.  In \cite{BSW02} we derived a
condition sufficient for the leading-order correction to the diagonal
approximation which was slightly stronger than Eq.~(\ref{eq:ergodic-tau}): The
speed of convergence to equidistribution with increasing $B$ cannot be
arbitrarily slow.  However, exponential convergence (corresponding to a
spectral gap of $M$ which is bounded away from zero uniformly in $B$) is
sufficient in any case.  We will restrict ourselves to graphs which obey this
condition rather than derive a more precise condition for the applicability of
Eq.~(\ref{eq:ergodic-tau}) to the summation of third-order diagrams.

A connection between the quantum form factor Eq.~(\ref{ff}) and the classical
dynamics given by Eq.~(\ref{matrixM}) can be established by representing the
form factor as a sum over (classical) POs.  We expand the matrix powers of $S$
in Eq.~(\ref{ff}) and obtain sums over products of matrix elements
$S_{p_2p_1,p_1p_t} \cdots S_{p_4p_3,p_3p_2}S_{p_3p_2,p_2p_1}$.  Obviously each
such product can be specified by a sequence of $t$ vertices.  Vertex sequences
which are identical up to a cyclic shift give identical contributions and will
be combined into the contribution of a periodic orbit $P=[p_{1},\dots,p_{t}]$.
For most POs there are $t$ different cyclic shifts.  Exceptions to this rule
are possible if a PO is a repetition of a shorter orbit, but the fraction of
such orbits decreases exponentially in $t$.  Moreover, if we assume the
existence of the limit (\ref{fflimit}), we can approach it through sequences
of prime $t$, which totally excludes repetitions.  We obtain
\begin{equation}\label{trace-formula}
  {\rm tr}S^{t}=t\sum_{P}A_{P}\e^{\i \phi_{P}}
\end{equation}
with $A_{P}=\prod_{i=1}^{t}\sigma^{(p_i)}_{p_{i+1},p_{i-1}}$ and
$\phi_{P}=\sum_{i=1}^{t}\phi_{p_{i+1},p_i}$ (vertex indices are taken modulo
$t$).  Substituting this into Eq.~(\ref{ff}) we obtain a double sum over
periodic orbits
\begin{equation}
  \label{ff-with<..>}
  K^{(B)}(\tau)={t^{2}\over B}\left\langle\sum_{P,Q}A_{P}A_{Q}^*
    \e^{\i(\phi_{P}-\phi_{Q})}\right\rangle_{\{\phi\}}\,.
\end{equation}
We can now perform the average over the phases $\phi_{ml}$ associated with the
directed bonds. If the system does not have time-reversal symmetry, all bond
phases can be varied independently.  The total phase of an orbit, $\phi_{P}$, 
can be written as linear combinations of the bond phases,
$\phi_{P}=\sum_{lm}n^{(P)}_{lm}\phi_{lm}$, where $n^{(P)}_{lm}$ counts visits
of the orbit $P$ to bond $(m,l)$.  Then we can average over
$\phi_{lm}$ using
\begin{eqnarray}
  \left< \e^{\i(n^{(P)}_{lm}\phi_{lm} - n^{(Q)}_{lm}\phi_{lm})} 
  \right>_{\phi_{lm}} = \de_{n^{(P)}_{lm},n^{(Q)}_{lm}}\,.
\end{eqnarray}
Thus averaging over all bond phases, $\{\phi\}$, amounts to picking out only
those pairs of orbits which visit the same set of bonds the same number of
times.  Therefore the form factor for a quantum graph with no time-reversal
symmetry (NTR) is
\begin{equation}
\label{eq:ff-PO-NTR}
K^{(B)}_{\rm NTR}(\tau)={t^{2}\over B}
\sum_{P,Q}A_{P}A_{Q}^*  
\bigg[\, {\prod_{lm}}\de_{n^{(P)}_{lm},n^{(Q)}_{lm}} \, \bigg]  \ .
\end{equation}

Time-reversal symmetry implies symmetric vertex-scattering matrices
\begin{equation}\label{symsig}
\sigma_{m'm}^{(l)}=\sigma_{mm'}^{(l)}
\end{equation}
and that the phases of a pair of bonds $(m,l)$ and $(l,m)$
related by time-reversal obey $\phi_{ml}=\phi_{lm}$.  The average in
Eq.~(\ref{ff-with<..>}) runs over all independent phases and results in
\begin{eqnarray}
\left< \e^{\i(\phi_P - \phi_Q)} \right>_{\{\phi\}}
= \prod_{l\geq m}\de_{n^{(P)}_{lm}+n^{(P)}_{ml},n^{(Q)}_{lm}+n^{(Q)}_{ml}}\,.
\end{eqnarray}
Thus averaging over all independent bond phases, $\{\phi\}$, amounts in this
case to picking out only those pairs of orbits which visit the same set of
bonds, or their time-reverses, the same number of times.  Hence, the form
factor for a quantum graph with time-reversal symmetry is
\begin{equation}
\label{eq:ff-PO-TR}
K^{(B)}_{\rm TR}(\tau)={t^{2}\over B}
\sum_{P,Q}A_{P}A_{Q}^*  
\bigg[ \, {\prod_{l\geq m}}
\de_{n^{(P)}_{lm}+n^{(P)}_{ml},n^{(Q)}_{lm}+n^{(Q)}_{ml}} \, \bigg]  \ .
\end{equation}
If the graph is defined to have bonds with fixed lengths and magnetic
vector potential as in \cite{KS97}, we can average over an
infinite energy window (for NTR-systems we also average over the
vector potential). Then the orbit pairs contributing to the form factor
are those where $P$, $Q$ have exactly the same lengths.
On a graph with rationally independent bond lengths this is equivalent to
\begin{eqnarray}
  \label{equallength_NTR}
  {\rm NTR:} \qquad \qquad \,  n^{(P)}_{lm}=n^{(Q)}_{lm}
  \qquad \qquad \ \  \forall l,m  \\
  \label{equallength_TR}
  {\rm TR:} \qquad   n^{(P)}_{lm}+n^{(P)}_{ml}=n^{(Q)}_{lm}+n^{(Q)}_{ml}
  \qquad\forall l,m  \,,
\end{eqnarray} 
so this averaging procedure also leads to Eqs.~(\ref{eq:ff-PO-NTR})
and (\ref{eq:ff-PO-TR}).


\section{The expansion in self-intersections of the periodic orbits}

\subsection{From orbits to diagrams}

\begin{figure}
  \def\fscale{0.60}
  \centerline{\includegraphics[scale=\fscale]{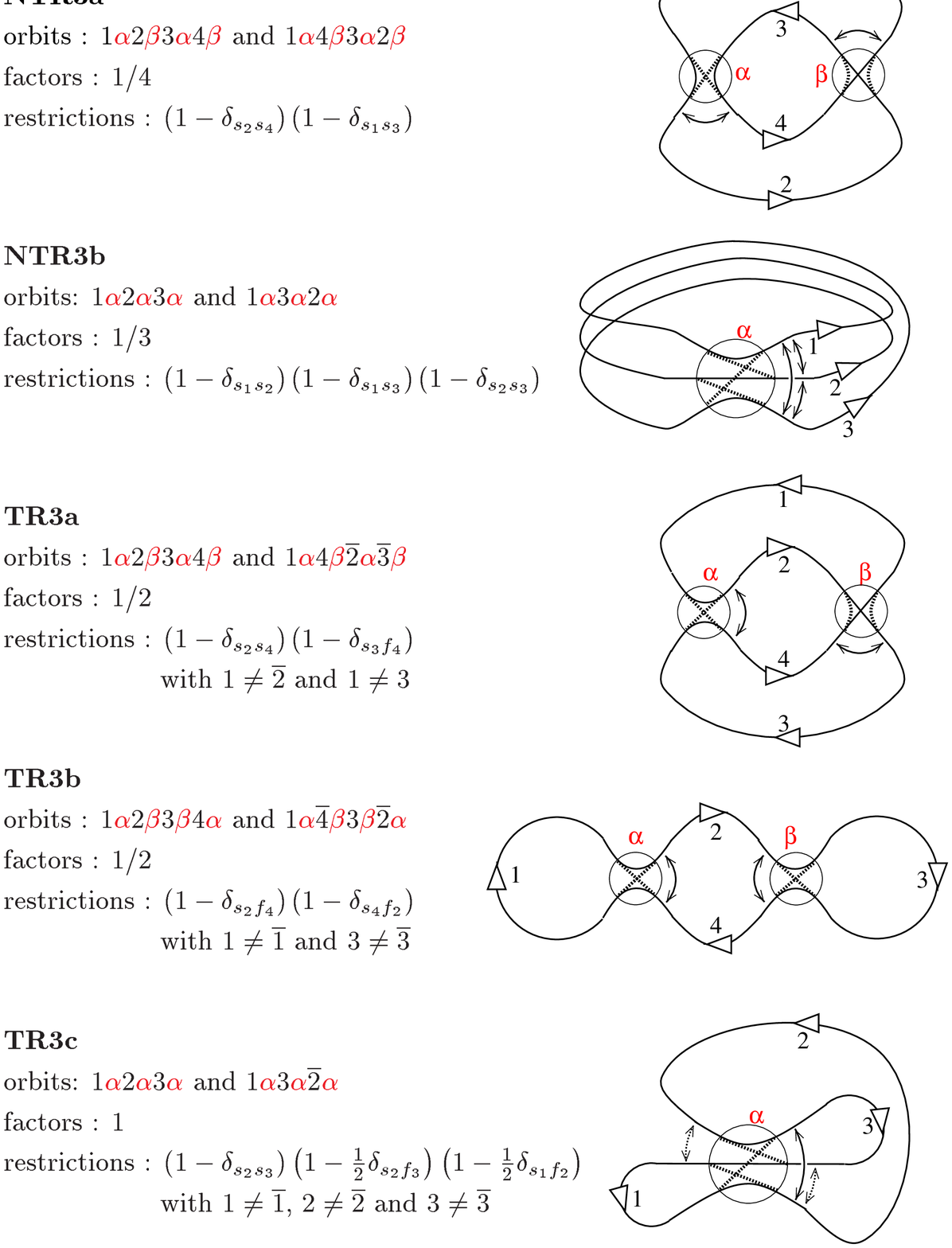}}
  \caption{
    Topology of NTR3a, NTR3b, TR3a, TR3b and TR3c.  In each case a pair of
    orbits is shown. One follows the solid line throughout (in the direction
    marked by triangular arrows). The second follows the solid lines (possibly
    with reversed direction) except at the intersections (denoted by circles)
    where it follows the dotted line.  Each circle represents a single vertex
    where a self-intersection of the orbit occurs.  Next to each topology we
    give the corresponding weight factor
    (Section~\protect\ref{sect:doublecountI}) and the restrictions
    (Section~\protect\ref{sect:doublecountII}).  Restrictions are indicated by
    the double-headed arrows.  Solid arrows indicate a ``full'' restriction of
    the form $\left( 1- \delta_{i j}\right)$, while dotted arrows indicate a
    ``half'' restriction of the form $\left( 1- \frac12 \delta_{i j}\right)$.}
  \label{fig1}
\end{figure}

The calculation of the form factor is now reduced to a combinatorial problem:
The sum over the pairs $P,Q$ in Eq.~(\ref{eq:ff-PO-NTR}) 
and Eq.~(\ref{eq:ff-PO-TR}) must be organized such that
Eqs.~(\ref{equallength_NTR}) or (\ref{equallength_TR}) 
are satisfied.  This can be done by composing $P$ and $Q$ from the same 
segments, or {\em arcs}, which appear in $P$
and $Q$ in different order and/or orientation.  This is possible if the orbit
$P$ contains self-intersections, i.e.\ vertices which are traversed more than
once, see Fig.~\ref{fig1} for examples. In general an orbit $P$ has many
self-intersections and many partner orbits $Q$ satisfying
Eqs.~(\ref{equallength_NTR}), (\ref{equallength_TR}) such that a summation
over all possible $Q$ for a fixed $P$ is too complicated. Instead we fix a
permutation of arcs followed by the time reversal of selected arcs and sum
first over all possible pairs of orbits $P,Q$ related by this transformation.
The clearest way to represent all possible transformations is graphical
(Fig.~\ref{fig1}), hence we refer to them as diagrams. The sum over all
diagrams finally gives the form factor.

The main problem with this approach is to ensure that each orbit pair $P,Q$ is
counted once and only once. This is difficult because for some pairs $P,Q$
the operation transforming $P$ into $Q$ is not unique.
Such orbit pairs are relatively rare in number but nevertheless they 
give essential contributions to the form factor \cite{BSW02}.  We will
explain our techniques for preventing the double counting of orbit pairs in
Sections~\ref{sect:doublecountI} and \ref{sect:doublecountII}.

If we consider $P$ as a single arc with no intersections, $Q=P$ is the only
possibility in the NTR case.  For TR the orientation of the arc can be
reversed such that $Q=\overline{P}$ is a second option.  The corresponding
diagrams have a simple circular shape.  Summation over these orbit pairs is
nothing other than the diagonal approximation. It produces $K_{\rm NTR1}=\tau$
and $K_{\rm TR1} = 2\tau$, respectively.  In \cite{BSW02} we considered 
orbits, $P$, made from two arcs, $1$ and $2$, joined at a single intersection
$\al$ and evaluated the (off-diagonal) contribution corresponding to
the resulting 
8-shaped diagram.  We found this gave rise to the second-order term in
Eq.~(\ref{kgoe}), $K_{\rm TR2} = -2\tau^{2}$, while there is no contribution of
this order for a NTR-system.  In this paper we calculate the
$\tau^3$-contribution by assuming that $P$ contains three or four arcs
connected at intersections.  A discussion of why only these
particular diagrams contribute to the $\tau^3$-contribution is deferred to
Section~\ref{sect:estimates}.

We begin by listing in Fig.~\ref{fig1} all
diagrams which contribute at third order in $\tau$.  
We denote arcs by numbers $1, 2, \ldots$ and the intersection points by Greek
letters $\alpha, \beta, \ldots$.  An arc can be identified by a sequence of
vertices, which does not include the intersection vertices, or, alternatively,
by a sequence of bonds, which includes the bonds from and to the intersection
points.  The length of the $i$th arc is denoted by $t_i$ and is defined as the
number of bonds in the arc (which is one more than the number of vertices in
the arc).  The sum of the lengths of all arcs gives $t$, the length of the
orbit.  The length of an arc is at least one.
For an arc $i$ leading from $\alpha$ to $\beta$ we denote the first vertex
following $\alpha$ by $s_i$ and the last vertex before $\beta$ by $f_i$.  In
the degenerate case when the arc going from $\alpha$ to $\beta$ is the single
bond $(\alpha,\beta)$ and does not contain any vertices ($t_i=1$) this implies
$s_i = \beta$ and $f_i = \alpha$.
  
As shown in Fig.~\ref{fig1}, the arcs forming an orbit $P$ and its
partner $Q$ are identical, but the way they are connected at the intersections
differs.  The orbit $P$ is given by the connections drawn as continuous lines,
while its partner orbit $Q$ is given by connections drawn as dotted lines.
The orbits $P$ and $Q$ are also written as a symbolic code to the left of
each diagram: a path that goes from the beginning of arc $1$ to vertex $\al$
then on arc $2$ to vertex $\be$ and so on is denoted as $1\al 2 \be \cdots$.
The diagrams in Fig.~\ref{fig1} divide into two classes, NTR and TR.  In the
NTR-diagrams all the arcs of $Q$ have the same orientation as the
corresponding arcs in $P$, while in the TR-diagrams some of the arcs of $Q$
are time-reversed.  For a system with no time-reversal symmetry, only the two
NTR-diagrams are possible, thus there are two
$\tau^3$-contributions to the form factor,
\begin{eqnarray}
  \label{eq:NTR3sum}
  K_{\rm NTR3} = K_{\rm NTR3a} + K_{\rm NTR3b}  \  .
\end{eqnarray}
For a system with time-reversal symmetry (TR), diagrams in both classes
contribute and the form factor is a sum of five terms
\begin{eqnarray}
  \label{eq:TR3sum}
  K_{\rm TR3} = 2 \left( 
    K_{\rm NTR3a}+K_{\rm NTR3b}+K_{\rm TR3a}+K_{\rm TR3b}+K_{\rm TR3c} 
  \right) \  .
\end{eqnarray}
The factor of two is due to the fact that for every diagram in
Fig.~\ref{fig1} there is another one with $Q$ replaced by its complete
time-reversal, $\overline{Q}$, which gives an identical contribution. 

\subsection{Avoiding double-counting I: multiplicity factors}

\label{sect:doublecountI}

The set of diagrams possesses certain degeneracies which can be
accounted for by simple prefactors multiplying the contributions.  One
such degeneracy is taken care of by the factor of two in
Eq.~(\ref{eq:TR3sum}).  In this subsection we discuss how to determine the 
other multiplicity factors arising due to the cyclicity of the POs and 
symmetries in the diagrams.  To sum over all orbit pairs $P,Q$ for a given 
diagram we sum over all possible arcs forming the orbit $P$.  
Consider the diagram NTR3a as an example.  Let $l_1$ and $l_3$ be some
fixed arcs starting at $\beta$ and ending at $\alpha$, while $l_2$ and $l_4$
denote arcs from $\alpha$ to $\beta$.  As we sum over all possible
realisations of arcs $1,2,3,4$ in NTR3a, we encounter a particular orbit $P$
where these arcs are given by
\begin{equation}
  \label{eq:assignmentP}
  1 = l_1,\quad 
  2 = l_2,\quad
  3 = l_3,\quad
  4 = l_4.\quad
\end{equation}
However we also encounter the orbit $P'$ where the arcs are
\begin{equation}
  \label{eq:assignmentP'}
  1 = l_3,\quad 
  2 = l_4,\quad
  3 = l_1,\quad
  4 = l_2.\quad
\end{equation}
The orbit $P'$ is related to $P$ by a cyclic shift and, therefore, it is
actually the same orbit.  As we are focusing on {\it pairs\/} of orbits, we
check the partner orbits resulting from $P$ and $P'$, too.  The partners for
$P$ and $P'$ are $Q = [l_1, \alpha, l_4, \beta, l_3, \alpha, l_2, \beta]$ and
$Q' = [l_3, \alpha, l_2, \beta, l_1, \alpha, l_4, \beta]$, respectively, and
they are also related by a cyclic shift.  Hence in the process of summation we 
will encounter the same orbit pair four times, once for each of the four
possible cyclic permutations of $P$.  To compensate for this we
introduce a multiplicity factor of a quarter.

To put this formally, we denote by $q(P)$ the operation transforming $P$ into
$Q$ for a given diagram, e.~g., $q_{\rm NTR3a}([1234])=[1432]$ and $q_{\rm
  TR3a}([1234])=[14\bar 2\bar 3]$. Further we denote by $\sigma$ the (cyclic)
left shift of the symbolic code, i.~e.\ $\sigma([1234])=[2341]$. To determine
the multiplicity factor we need to count all cyclic permutations $\sigma^k$
such that
\begin{equation}\label{sigk}
q\circ\sigma^{k}(P) 
=
\sigma^{k'}\circ q(P)
\qquad\mbox{or}\qquad
{q\circ \sigma^{k}(P)}
=
\sigma^{k'}\circ \overline{q(P)}
\end{equation}
for some $k'$
and arbitrary $P$, i.~e.\ application of $q$ to the shifted code yields the
same or the completely time-reversed result, up to another shift.  Obviously
the second option is only applicable to the TR diagrams.  Noting that
the trivial solution $k=k'=0$ is always available, we proceed to list
the factors for each diagram. 
\begin{itemize}
\item{NTR3a:} We have $q_{\rm NTR3a}\circ \sigma^{k}=\sigma^{4-k}\circ q_{\rm NTR3a}$ 
for any $k=0,1,2,3$ and consequently $m_{\rm NTR3a}=4$. 
\item{NTR3b:} Similarly we have $q_{\rm NTR3b}\circ \sigma^{k}=
\sigma^{3-k}\circ q_{\rm NTR3b}$ for $k=0,1,2$ and $m_{\rm NTR3b}=3$.
\item{TR3a:} The only nontrivial solution to Eq.~(\ref{sigk}) is
$k=2$ where $q_{\rm TR3a}\circ \sigma^{2}=\overline{q_{\rm TR3a}}$. 
Therefore we have $m_{\rm TR3a}=2$.
\item{TR3b:} The only nontrivial solution is $k=2$ where
  $q_{\rm TR3b}\circ\,\sigma^{2}=\sigma^{2}\circ{q_{\rm TR3b}}$. Therefore we
  have $m_{\rm TR3b}=2$.
\item{TR3c:} Eq.~(\ref{sigk}) has no solution besides the identity $k=0$,
  therefore $m_{\rm TR3c}=1$.
\end{itemize}
When we evaluate the contribution from each diagram we will include a factor
$1/m$ in order to compensate for the ambiguity just described.


\subsection{Avoiding double-counting II : restrictions and exceptions}
\label{sect:doublecountII}

\begin{figure}
  \def\fscale{0.7}
  \centerline{\includegraphics[scale=\fscale]{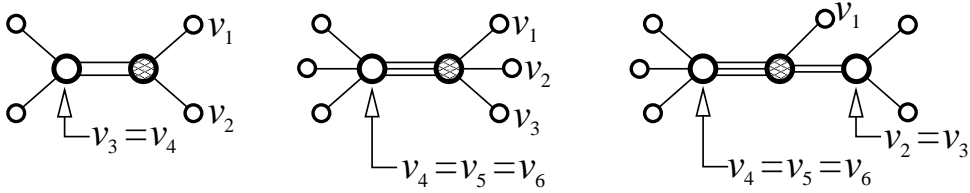}} 
  \caption{ Examples of ambiguous intersections, where we have removed the
  ambiguity by placing the intersection (the shaded vertex) as far to the 
  right as possible.  This is enforced in the left-hand diagram
  by introducing a factor of $\left( 1-\delta_{v_1v_2} \right)$, 
  while in the other diagrams it is enforced by a factor of 
  $\left( 1-\delta_{v_1v_2}\delta_{v_1v_3} \right)$. }
  \label{fig2}
\end{figure}

As shown in Ref.~\cite{BSW02}, {\it tangential} self-intersections of orbits
are a potential source for double-counting of orbits which must carefully be
avoided.  By a {\it tangential} intersection we mean the situation where an
orbit does not merely cross itself but follows itself for at least one bond.
For example the orbit
\begin{equation}
  \cdots \to f_{i} \to \alpha \to \beta \to s_{i+1} \to \cdots \to f_{j+1} 
  \to \beta \to \alpha  \to s_{j+1} \to \cdots
\end{equation}
crosses itself along the non-directed bond $(\alpha,\beta)$.  It is easy to
mistakenly count such an orbit once with $\alpha$ as the intersection point
and once with $\beta$ as the intersection point.  We avoid this using a method
outlined in Ref.~\cite{BSW02}.  We uniquely define the intersection point by
ruling that if there is an ambiguity then the intersection is as far to one
side as possible.  As an example we show some ambiguous intersections in
Fig.~\ref{fig2} and insist the intersection is as far to the right as
possible.  For the 2-intersection we do this by demanding that $v_1 \neq v_2$,
this is achieved by introducing a factor of the type $\left(
  1-\delta_{v_1v_2}\right)$, referred to as a {\it restriction} on the
diagram. For the 3-intersections a restriction of the form $\left(
  1-\delta_{v_1v_2}\delta_{v_1v_3}\right)$ removes the ambiguity.  However, we
will not actually use this restriction on any 3-intersection, because we will
see that {\em stronger} restrictions apply in the diagrams we evaluate.

For NTR3a, TR3a and TR3b we choose the following restrictions to ensure the
ambiguities at intersections are removed
\begin{itemize} 
\item{NTR3a:} 
  $\Delta_{\rm NTR3a}
  =\left( 1-\delta_{s_2s_4} \right) \left( 1-\delta_{s_1s_3} \right)$
\item{TR3a:}
  $\Delta_{\rm TR3a}
  =\left( 1-\delta_{s_2s_4} \right) \left( 1-\delta_{s_3f_4} \right)$
\item{TR3b:}
  $\Delta_{\rm TR3b}
  =\left( 1-\delta_{s_2f_4} \right) \left( 1-\delta_{s_4f_2} \right)$,
\end{itemize}
where $s_i$ denotes the first vertex on the arc $i$ and $f_i$ denotes the
last.  We wish to emphasize that there is no unique way of imposing the
restrictions, since they are merely convenient ways of excluding the
double counting of certain contributions.  What is more, the individual results for
NTR3a, NTR3b, TR3a, TR3b and TR3c may depend on the particular choice of
restrictions.  Only the final sums in Eq.~(\ref{eq:NTR3sum}) and
(\ref{eq:TR3sum}) do not depend on them.

Now that we come to NTR3b, we will see exactly how much freedom we have in
choosing restrictions.  First we want to ensure that we count tangential
intersections correctly.  For a 3-intersection, such as the one in
NTR3b, we could do this by setting $\Delta_{\rm 
  NTR3b} =\left( 1-\delta_{s_1s_2}\delta_{s_2s_3} \right)$.  However
we also notice that there are ambiguous contributions which could be
counted in either NTR3a or NTR3b
\footnote[2]{This abundance of choice, when any
    of the three NTR3b diagrams is equivalent to any of the four NTR3a
    diagram is another manifestation of the cyclic symmetry
    discussed in Section~\ref{sect:doublecountI} and is taken care of
    by the multiplicity factors.}:
\begin{itemize} 
\item NTR3b with either $s_1=s_2$, $s_2=s_3$ or $s_1=s_3$ is
  equivalent to NTR3a with any of the following: ($t_1=1$ \&
  $f_1=f_3$), ($t_2=1$ \& $f_2=f_4$), ($t_3=1$ \& $f_1=f_3$) or
  ($t_4=1$ \& $f_2=f_4$)
\item NTR3b with either $f_1=f_2$, $f_2=f_3$ or $f_1=f_3$ is
  equivalent to NTR3a with any of the following: 
  ($t_1=1$ \& $s_1=s_3$), ($t_2=1$ \& $s_2=s_4$), ($t_3=1$ \&
  $s_1=s_3$) or ($t_4=1$ \& $s_2=s_4$).
\end{itemize} 
Obviously we should only count each of these contributions once, but we have
the freedom to choose whether we count each of them in NTR3a or NTR3b.  The
physical quantities (\ref{eq:NTR3sum}) and (\ref{eq:TR3sum}) contain the sum
of NTR3a and NTR3b, so all choices are strictly equivalent.  
However given that we have imposed the restriction $s_1\neq s_3$ on NTR3a 
the second type of orbits (NTR3b with $f_i=f_j$) cannot belong to NTR3a.  
Thus, once we have
chosen the above restrictions for NTR3a we are forced into the choice
\begin{eqnarray}
  \label{eq:clvr-restriction}
  \Delta_{\rm NTR3b}
  =\left( 1-\delta_{s_1s_2}\right) 
  \left( 1-\delta_{s_1s_3}\right) 
  \left( 1-\delta_{s_2s_3}\right).
\end{eqnarray}

Before we can move on to TR3c, we must first look carefully at the restriction
we placed on TR3a.  In Section~\ref{sect:doublecountI} we introduced the
factor of $1/2$ to avoid double-counting.  The double-counting in this
particular instance was caused by the permutation $\sigma^2=[3412]$ which
swaps around arcs $1 \leftrightarrow 3$ and $2 \leftrightarrow 4$ and produces
a pair $P'=\sigma^2(P)$ and $Q'=\sigma^2(Q)$, which is identical to $P,Q$ up
to a shift.  However the restriction $s_2\neq f_4$ that we introduced on TR3a
is {\it not\/} symmetric with respect to this permutation.  
For the orbits satisfying
$s_1\ne f_2$ and $s_3 \ne f_4$ this does not present any problems.  Let us
consider what happens when arcs $1$ and $2$ are different but have $s_1 =
f_2$.  This orbit is still counted twice in the summation over all possible
arcs, but in the second instance the intersection point $\beta$ is shifted,
resulting in $t_1' = t_3+1$, $t_2' = t_4+1$, $t_3' = t_1-1$, $t_4' = t_2-1$.
We illustrate that by the following example of orbits which contribute to TR3a. 
The pair
\begin{equation}
  \label{eq:dodgy_TR3a}
  P = [\beta,\gamma,a,\alpha,b,\gamma,\beta,d,\alpha,c]
  \quad\mbox{and}\quad
  Q = [\beta,\gamma,a,\alpha,c,\beta,\gamma,b,\alpha,d]
\end{equation}
is obtained by combining the arcs
\begin{eqnarray}
  \label{eq:assig1}
  \hbox{arc }1 &=& (\beta, \gamma) \ra (\gamma, a) \ra (a, \alpha)
  \qquad \qquad \hbox{arc }3 = (\beta, d) \ra (d, \alpha)\\
  \hbox{arc }2 &=& (\alpha, b) \ra (b, \gamma) \ra (\gamma, \beta)
  \qquad \qquad \hbox{arc }4 = (\alpha, c) \ra (c, \beta)
\end{eqnarray}
with the intersection points $\alpha$ and $\beta$, or by combining the
arcs
\begin{eqnarray}
  \label{eq:assig2}
  \hbox{arc }1 &=& (\gamma, \beta) \ra (\beta, d) \ra (d, \alpha)
  \qquad \qquad \hbox{arc }3 = (\gamma, a) \ra (a, \alpha)\\
  \hbox{arc }2 &=& (\alpha, c) \ra (c, \beta) \ra (\beta, \gamma)
  \qquad \qquad \hbox{arc }4 = (\alpha, b) \ra (b, \gamma)
\end{eqnarray}
with the intersection points $\alpha$ and $\gamma$.  We therefore see that the
factor of $1/2$ works also when $s_1 = f_2$.  But not when, in addition to
$s_1 = f_2$, $t_1$ or $t_2$ is equal to 1.  These orbits appear in the sum for
TR3a only once and are subsequently multiplied by $1/2$, so it appears we
miscount their contribution. On the other hand these orbits can also be
counted in TR3c, as shown below.  We find it convenient to keep the above
restriction on TR3a, thus counting half their contribution in TR3a.  This
forces us to count the other half of their contribution in TR3c.


Now we can move on to finding the restrictions on TR3c.
First we list the special cases of TR3c which could be counted in other 
diagrams.
\begin{enumerate}
\item 
  \label{item:s1f2}
  TR3c with $s_1=f_2$ is equiv. to TR3a with ($t_3=1$ \&
  $s_3=f_4$) or ($t_1=1$ \& $s_1=f_2$)
\item 
  \label{item:s2f3}
  TR3c with $s_2=f_3$ is equiv. to TR3a with ($t_4=1$ \&
  $s_3=f_4$) or ($t_2=1$ \& $s_1=f_2$)
  
\item 
  \label{item:s2s3}
  TR3c with $s_2=s_3$ is equiv. to TR3a with ($t_1=1$ \&
  $f_1=f_3$) or ($t_3=1$ \& $f_1=f_3$)
\item 
  \label{item:f1f2}
  TR3c with $f_1=f_2$ is equiv. to TR3a with ($t_2=1$ \&
  $s_2=s_4$) or ($t_4=1$ \& $s_2=s_4$)
  
\item 
  \label{item:s1f1}
  TR3c with $s_1=f_1$ is equiv. to TR3b with ($t_2=1$ \&
  $s_2=f_4$) or ($t_4=1$ \& $s_4=f_2$)
\item 
  \label{item:s3f3}
  TR3c with $s_3=f_3$ is equiv. to TR3b with ($t_2=1$ \&
  $s_2=f_4$) or ($t_4=1$ \& $s_4=f_2$)
\end{enumerate}
Now we carefully count in TR3c only those contributions which have not
already been counted in TR3a or TR3b.  Lines (\ref{item:s1f2}) and
(\ref{item:s2f3}) above show that cases $s_1=f_2$ and $s_2=f_3$ should
be counted in TR3c with the factor $1/2$.  Line (\ref{item:s2s3})
shows that the case $s_2=s_3$ should not be counted in TR3c as it is
fully counted in TR3a; the case $f_1=f_2$, line (\ref{item:f1f2}),
should be fully counted in TR3c.  Lines (\ref{item:s1f1}) and
(\ref{item:s3f3}) show that the cases $s_1=f_1$ and $s_3=f_3$ are not
counted in TR3b and should be fully counted in TR3c.
All this is realised by the restrictions
\begin{eqnarray}
  \Delta_{\rm TR3c}=
  \left( 1-\delta_{s_2s_3}\right) \left( 1-\half\delta_{s_2f_3}\right)
  \left( 1-\half\delta_{s_1f_2}\right)\,.
\end{eqnarray}

Above we have ensured that no orbits are double-counted among the
diagrams NTR3a, NTR3b, TR3a, TR3b and TR3c.  However we should also
exclude the orbits that have already been counted at lower orders of
the expansion.
Considering NTR3a, if arc 1 is identical to arc 3 (or arc 2 identical
to arc 4), the diagram is reduced to giving a contribution to the
diagonal approximation, so it should not be counted here.
Fortunately, the restrictions we have put on NTR3a ensure that this
contribution is not counted.  Moving on to NTR3b, if any two arcs in
the NTR3b diagram are self-retracing the diagram reduces to a diagram 
already counted as a $\tau^2$-contribution in a TR-system.  
Therefore, in the TR case, we should subtract
its contribution from the sum.  However, we will see at the end of
Section~\ref{sect:sum_NTR3} that such a contribution is zero.

For TR3a, we insist that $1 \ne \bar{2}$, $1 \ne 3$, $4 \ne 2$ and $4
\ne \bar{3}$ because the orbits breaking these rules have already been
counted at ${\cal O}[\tau^2]$ of the expansion.  For the same reason
we insist that TR3b obeys $1 \ne \bar{1}$, $3 \ne \bar{3}$, $2 \ne
\bar{4}$, while TR3c obeys $1 \ne \bar{1}$, $2 \ne \bar{2}$ and $3 \ne
\bar{3}$.  Note that some of the restrictions are superfluous since
they refer to orbits that are already excluded.  For example we can
drop the restriction $4 \ne \bar{3}$ because this is automatically
enforced by the stronger restriction $s_3\neq f_4$.

The complete set of restrictions is the following
\begin{itemize}
\item{NTR3a:}   
   $\De_{\rm NTR3a}=
   \left( 1-\delta_{s_2s_4}\right)\left( 1-\delta_{s_1s_3}\right)$ 
\item{NTR3b:}  
   $\De_{\rm NTR3b}=
   \left( 1-\delta_{s_1s_2}\right) \left( 1-\delta_{s_1s_3}\right) 
   \left( 1-\delta_{s_2s_3}\right) $ 
   where orbits with $(2,3) = (\bar{2},\bar{3})$ must be subtracted for
   systems with TR symmetry.
\item{TR3a:}
   $\De_{\rm TR3a}= 
   \left( 1-\delta_{s_2s_4}\right)\left( 1-\delta_{s_3f_4}\right)$ 
   with $1 \ne \bar{2}$ and  $1 \ne 3$.
\item{TR3b:}
   $\De_{\rm TR3b}=
   \left( 1-\delta_{s_2f_4}\right)\left( 1-\delta_{s_4f_2}\right)$
   with $1 \ne \bar{1}$ and $3 \ne \bar{3}$.
\item{TR3c:}
   $\De_{\rm TR3c}=
   \left( 1-\delta_{s_2s_3}\right)\left( 1-\frac12\delta_{s_2f_3}\right)
   \left( 1-\frac12\delta_{s_1f_2}\right)$
   with $1 \ne \bar{1}$, $2 \ne \bar{2}$ and $3 \ne \bar{3}$.
\end{itemize}
We reiterate that this self-consistent set of restrictions is not unique.
And, although this choice lead to simpler calculations than all the others we
tried, we can not rule out the possibility that there is another
self-consistent set of restriction which would further simplify our
calculations.


\subsection{Orbit amplitudes}

Before we can attempt the summation over all orbit pairs $P,Q$ within a given
diagram, we still need to understand the structure of the product $A_P
A^*_Q$ appearing in Eqs.~(\ref{eq:ff-PO-NTR}), (\ref{eq:ff-PO-TR}).  We
consider the diagram NTR3b as an example.  Let arc 1 be of length $t_1$,
consisting of the vertices $[x_1, x_2, \ldots x_{t_1-1}]$, where 
$x_1\equiv s_1$ and $x_{t_1-1}\equiv f_1$.
Then both $A_P$ and $A_Q$ will contain factors
$\sigma^{(x_1)}_{x_2,\alpha}$, $\sigma^{(x_2)}_{x_3,x_1}$,
$\sigma^{(x_3)}_{x_4,x_2}, \dots, \sigma^{(x_{t_1-1})}_{\alpha,x_{t_1-2}}$ .  
Thus when we evaluate the product $A_P
A^*_Q$, the contribution of the arc 1 will come in the form
\begin{equation}
  \label{eq:arc1_contrib}
  |\sigma^{(x_1)}_{x_2,\alpha} \sigma^{(x_2)}_{x_3,x_1}
  \sigma^{(x_3)}_{x_4,x_2} \cdots \sigma^{(x_{t_1-1})}_{\alpha,x_{t_1-2}} |^2 
  = P_{ (\alpha, x_1) \to (x_1, x_2) \to \ldots \to (x_{t_1-1}, \alpha)}  
  \equiv P_1 \ ,
\end{equation}
which is the {\it classical} probability of following arc 1 from bond 
$(\alpha,s_1)$ to bond $(f_1,\alpha)$\footnote{
$P_{1}=1$ if arc 1 contains no vertices, i.~e.\ if  $t_1=1$.}.  
Analogous considerations for arcs 2 and 3 lead to the
probabilities $P_2$ and $P_3$. The factors not yet accounted for in
$P_1,P_2,P_3$ are the transition amplitudes picked up at the intersection
vertex $\alpha$:
\begin{equation}
  \label{eq:productA_PA_Q}
  A_P A^*_Q = P_1 \times P_2 \times P_3 \times
  \sigma^{(\alpha)}_{s_3f_2} 
  \sigma^{(\alpha)}_{s_2f_1}
  \sigma^{(\alpha)}_{s_1f_3} 
\times
  \Big(
\sigma^{(\alpha)}_{s_2f_3} 
\sigma^{(\alpha)}_{s_3f_1}
\sigma^{(\alpha)}_{s_1f_2}
\Big)^*.
\end{equation}
To evaluate the contribution of a given diagram a product like
Eq.~(\ref{eq:productA_PA_Q}) must be summed over all free parameters, namely
all intersection points and all possible arcs connecting these points. The
latter summation includes a sum over the lengths $t_i$ of these arcs with the
restriction that the total length of the orbit is $t$.  

The summation over all the intermediate vertices $x_2, x_3, \ldots, x_{t_1-2}$
along arc 1 can be performed immediately, because it is unaffected by the
restrictions discussed in the previous subsection.  This summation adds the
classical probabilities of all possible paths leading from bond $(\alpha,s_1)$
to bond $(f_1,\alpha)$ in $t_1-1$ steps and results consequently in the
classical transition probability $P_{(\alpha,s_1) \to (f_1,\alpha)}^{(t_1-1)}$
given by Eq.~(\ref{classtransprob}).  Analogous summations over the other arcs
produce $P_{(\alpha,s_2) \to (f_2,\alpha)}^{(t_2-1)}$ and 
$P_{(\alpha,s_3) \to (f_3,\alpha)}^{(t_3-1)}$.
The above approach extends trivially to the TR diagrams when we recall that
time-reversal symmetry implies that the matrices $\sigma^{(v)}$ are symmetric.

The remaining summation is over the lengths $t_i$ of all arcs, the first and
the last vertex $s_i$ and $f_i$ of all arcs $i$ with $t_i>1$ and the
intersection points like $\alpha$. For general graphs this sum is still too
complicated for explicit calculations, mainly because transition probabilities
like $P_{(\alpha,s_1) \to (f_1,\alpha)}^{(t_1-1)}$ are dependent on
the details of the topology of the graph.  For sufficiently long arcs,
however, these transition probabilities can be replaced by $B^{-1}$ according
to Eq.~(\ref{eq:ergodic-tau}). Then the sum over vertices decouples into a
product of sums associated with each self-intersection vertex $\alpha$ which
can finally be evaluated using the unitarity of the vertex-scattering matrices
$\sigma^{(\alpha)}$. This is the strategy we shall follow in the next two
sections, where explicit summation of the NTR3 and TR3 diagrams is performed.


\section{Summing the NTR contributions}

\subsection{Summation of NTR3 diagrams}
\label{sect:sum_NTR3}

Starting with the NTR3a diagram, we write
\begin{equation}
  \label{eq:NTR3a}
  \fl K_{\rm NTR3a}(\tau) = \frac{1}{4}\frac{t^2}{B}
  \sum_{\{t_i\}}\de \left[ t- {\textstyle \sum_{i=1}^4 t_i} \right]
  \sum_{\alpha,\beta} \sum_{\{s_i,f_i\}} 
  \Sigma_{\rm NTR3a} \times P_{\rm NTR3a} 
  \times \Delta_{\rm NTR3a} \ ,
\end{equation}
where 
\begin{eqnarray}
  \label{eq:components_NTR3a}
  \Sigma_{\rm NTR3a} 
  &=& 
\sigma^{(\alpha)}_{s_4f_3}
\sigma^{(\beta)}_{s_3f_2}
\sigma^{(\alpha)}_{s_2f_1}
\sigma^{(\beta)}_{s_1f_4}
\,
\sigma^{(\alpha)*}_{s_2f_3}
\sigma^{(\beta)*}_{s_3f_4}
\sigma^{(\alpha)*}_{s_4f_1}
\sigma^{(\beta)*}_{s_1f_2}
\\
  P_{\rm NTR3a} 
  &=& P_{(\beta,s_1) \ra (f_1,\alpha)}^{(t_1-1)}
  P_{(\alpha,s_2) \ra (f_2,\beta)}^{(t_2-1)}
  P_{(\beta,s_3) \ra (f_3,\alpha)}^{(t_3-1)}
  P_{(\alpha,s_4) \ra (f_4,\beta)}^{(t_4-1)}\\
  \Delta_{\rm NTR3a} 
  &=& (1-\delta_{s_1s_3})(1-\delta_{s_2s_4}).
\end{eqnarray}
As $t\to\infty$ and $t=t_1+t_2+t_3+t_4$, at least one of the arcs must be
long.  Without loss of generality we assume that $t_1\geq t/4$.  From
Eq.~(\ref{eq:ergodic-tau}) we have $P_{(\beta,s_1) \ra (f_1,\alpha)}^{(t_1-1)}
\approx B^{-1}$ and the only factors in Eq.~(\ref{eq:NTR3a})
depending on $f_1$ are 
$\sigma^{(\alpha)}_{s_2f_1} \sigma^{(\alpha)*}_{s_4f_1}$.
Using the unitarity of the $\sigma$-matrices we perform the summation
\begin{equation}
  \label{eq:unit_sigma}
  \sum_{f_1} \sigma^{(\alpha)}_{s_2f_1} \sigma^{(\alpha)*}_{s_4f_1}
  = \delta_{s_2s_4}.
\end{equation}
However the restriction $\De_{\rm NTR3a}$ contains the term 
$(1-\delta_{s_2s_4})$, leading to the result
\begin{eqnarray} 
  K_{\rm NTR3a} = 0 \ .
\end{eqnarray}
Calculation of $K_{\rm NTR3b}$ goes along the same route with
\begin{equation}
  \label{eq:NTR3b}
  \fl K_{\rm NTR3b}(\tau) = \frac{1}{3}\frac{t^2}{B}
  \sum_{\{t_i\}} \de\left[ t-{\textstyle \sum_{i=1}^3 t_i} \right]
  \sum_{\alpha} \sum_{\{s_i,f_i\}} 
  \Sigma_{\rm NTR3b} \times P_{\rm NTR3b} 
  \times \Delta_{\rm NTR3b} \ ,
\end{equation}
where 
\begin{eqnarray}
  \label{eq:components}
  \Sigma_{\rm NTR3b} &=& 
\sigma^{(\alpha)}_{s_3f_2}
\sigma^{(\alpha)}_{s_2f_1}
\sigma^{(\alpha)}_{s_1f_3}
\,
\sigma^{(\alpha)*}_{s_2f_3}
\sigma^{(\alpha)*}_{s_3f_1}
\sigma^{(\alpha)*}_{s_1f_2}
\\
  P_{\rm NTR3b} 
  &=& P_{(\alpha,s_1) \ra (f_1,\alpha)}^{(t_1-1)}
  P_{(\alpha,s_2) \ra (f_2,\alpha)}^{(t_2-1)}
  P_{(\alpha,s_3) \ra (f_3,\alpha)}^{(t_3-1)}\\
  \Delta_{\rm NTR3b} 
  &=& (1-\delta_{s_1s_2})(1-\delta_{s_2s_3})(1-\delta_{s_3s_1}).
\end{eqnarray}
Exactly as for NT3a we can sum over $f_i$ where arc $i$ is long,
which results in $\de$-function which we combine with $\Delta_{\rm NTR3b}$
to get the answer
\begin{eqnarray}
K_{\rm NTR3b} = 0 \,.  
\end{eqnarray}
The sum of the NTR3a and NTR3b diagrams vanishes and so
\begin{equation}\label{kntr}
K_{\rm NTR}(\tau) = 0 \,.
\end{equation}
Thus we see that for a wide class of quantum graphs without time-reversal
symmetry the $\tau^3$-contribution to the form factor is zero,
as expected from the BGS conjecture.

We wish to note that the derivation given above is relatively simple, since
NTR3a and NTR3b both vanish due to the particular choice of restrictions which
make orbit pairs in the intersection of NTR3a and NTR3b unique (see the
discussion near Eq.~(\ref{eq:clvr-restriction})).  
Had we chosen to assign all
ambiguous diagrams to NTR3a, then the results for NTR3a and NTR3b would both
have been non-zero, although the total sum $K_{\rm NTR}(\tau)$ would of \
course still have equalled zero.

To apply the above result to the TR case, described by (\ref{eq:TR3sum}), we
must subtract the contribution of the NTR3b diagram with two self-retracing
arcs, as discussed in Section~\ref{sect:doublecountII}.  We use the fact that 
long self-retracing arcs give only exponentially small corrections because the
number of free summation variables is reduced by a factor of two. 
Thus we only need to consider short self-retracing arcs. 
Without loss of generality we can
assume that arcs 2 and 3 are self-retracing and short, implying that $t_1$
must be long enough for $P^{(t_1-1)}=B^{-1}$ to hold.  Then the sum over $f_1$
results in a factor of $\de_{s_2s_3}$ and, combining this with the restriction
$\De_{\rm NTR3b}$, we find that the contribution of this case is identically
zero.  Thus the NTR diagrams contribute nothing to the form factor of
TR-systems.

\subsection{Generalization to higher orders}
\label{ntr:gen}

\begin{figure}
  \def\fscale{0.5}
  \centerline{\includegraphics[scale=\fscale]{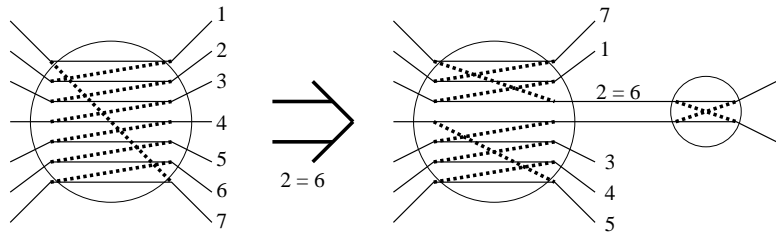}} 
  \caption{The picture on the left is a self-intersection in a
    NTR-contribution with seven crossing arcs.  If two bonds leaving this
    intersection coincide, in this case $2$ and $6$, the intersection can be
    redrawn (as on the right) as more than one intersections.  
    In this case there are three intersections, one 2-intersection,
    one 3-intersection and one 4-intersection, the latter two being
    at the same vertex. 
  \label{fig3}}
\end{figure}

One may speculate that the arguments given in the previous subsection
admit a straightforward generalization to higher-order diagrams.
Given an $n$th-order diagram, we impose the following restriction on
each of its intersections
\begin{equation}
  \label{eq:gen_restrictions}
  \De
  = \prod_{i,j} \left( 1- \de_{s_i s_j}\right)\ ,
\end{equation}
where the product is over the set of all arcs leaving the
intersection.  Now we can evaluate diagrams in the same way as we did for
$n=3$.  As soon as $n \ll t/t_{\rm erg}$ at least one arc must be long.
If arc $i$ is long then $P^{(t_i-1)} \simeq B^{-1}$ and the sum over
$f_i$ generates a $\de$-function.  Combining this $\de$-function with
the restriction at the vertex produces zero.

To justify the choice of the restriction (\ref{eq:gen_restrictions})
for any intersection, we notice that if any two bonds leaving the
vertex are the same, the intersection can be rearranged as a group of
more than one intersection, each satisfying the above restriction.
An example of such rearrangement is presented in Fig.~\ref{fig3}.  If
the original intersection was part of an $n$th-order diagram, then the
rearranged one is part of another valid $n$th-order diagram (as can be
shown by counting the powers of $B$, see Section~\ref{sect:estimates}).
The above restriction thereby helps to prevent double counting of
orbits with tangential intersections.

This argument essentially shows that the contribution of all
$n$th-order diagrams is zero in the NTR case.  However, an important
detail is missing: one has to show that all eligible pairs of periodic
orbits belong to {\it one and only one} diagram, i.e. that we did not
miss anything and did not count anything more than once.
Unfortunately we found pairs of orbits that violate both parts of this
statement.  These counterexamples seem to be ``rare'', in the sense
that the sum of their contributions vanish as $B\to\infty$, 
however a rigourous proof of this observation remains an open problem.

To summarise, a generalization of the argument of
Section~\ref{sect:sum_NTR3} sketches a proof of exactness of the
diagonal approximation for $\tau \le 1$ in the absence of
time-reversal symmetry.  To complete the proof one would have to
verify that the above restriction counts all relevant pairs of
orbits once and only once.


\section{Summation of TR3 diagrams for a fully-connected ``Fourier'' graph}
\label{sect:sum_TR3}

\begin{figure}
  \def\fscale{0.5}
  \centerline{\includegraphics[scale=\fscale]{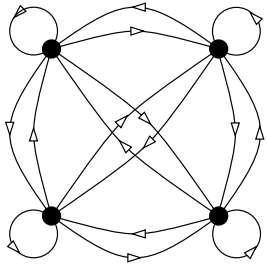}} 
  \caption{A fully-connected graph with four vertices and 
    sixteen directed bonds.}
  \label{fig4}
\end{figure}

Evaluating $K_{\rm TR3}$ for a general class of graphs is a
complicated and tedious task \cite{Bpre}.  Fortunately, the calculation
simplifies considerably for a special case described below.  In this
section we restrict our attention to fully-connected graphs with $N$
vertices and $B=N^{2}$ directed bonds, including a loop at each of the
vertices.  An example with $N=4$ is shown in Fig.~\ref{fig4}.  We
assume that the vertex-scattering matrices are
\begin{equation}\label{sigma}
  \sigma^{(l)}_{m'm}={1\over \sqrt{N}}\;\exp\Bigg(2\pi\i{mm'\over N}\Bigg)
\end{equation}
for all $l$.  These matrices were proposed in \cite{Tan01} and result in a
particularly fast convergence to RMT-like statistics.  Because of the analogy
to the discrete Fourier transformation from $m$ to $m'$, we call a vertex
endowed with the scattering matrix (\ref{sigma}) a ``Fourier''-vertex.  The
corresponding matrix $M$ represents uniform scattering at the vertex,
\begin{eqnarray}
M_{m'l,lm} = |\sigma^{(l)}_{m'm}|^2 = N^{-1} \ ,
\end{eqnarray}
and thus the probability to get from $(a, b)$ to $(c, d)$ in $t$
step is 
\begin{eqnarray}\label{clastransprob}
  P_{(a,b) \to (c,d)}^{(t)} 
  =\(M^{t}\)_{(d,c),(b,a)} 
  = \left\{ 
  \begin{array}{lc}
    N^{-1}\de_{b c}, & t=1\\ 
    B^{-1}, & t>1\,.
  \end{array} 
  \right.
\end{eqnarray}
It is also useful to have an expression for $\widetilde{P}_{(a,b) \to
  (b,a)}^{(t)}$, the probability to get from $(a, b)$ to $(b, a)$
following only self-retracing paths.  The contribution of each path is
$N^{-t}$ and, due to their special structure, a self-retracing path of
$2m+1$ or $2m+2$ steps will contain $m$ vertices not including the
initial $a$ and $b$.  Each of these $m$ vertices can be freely chosen from
the $N$ vertices of the graph, resulting in $N^m$ different
self-retracing paths.  Thus, the probability $\widetilde{P}_{(a,b) \to
  (b,a)}^{(t)}$ takes the form
\begin{equation}\label{Ptilde}
  \widetilde{P}_{(a,b) \to (b,a)}^{(t)} 
  = N^{-1-m}, \qquad \mbox{with } t = 2m\mbox{ or }2m+1\,,
\label{eq:self-retracing}
\end{equation}
i.~e.\  it is indeed decaying exponentially in time.


\subsection{Summation of TR3a}

Here we calculate the contributions of orbits with the topology of TR3a 
which obey the conditions $s_2 \ne s_4$ and $s_3 \ne f_4$.  
We enforce these conditions by multiplying the contribution of all orbits of 
this topology by 
\begin{eqnarray}
\Delta_{\rm TR3a} =(1-\de_{s_2 s_4})(1-\de_{s_3 f_4}) .
\end{eqnarray}
Thus the contribution of TR3a is 
\begin{equation}
  \label{eq:fin_TR3a}
  K_{\rm TR3a}(\tau) = \frac12 \frac{t^2}{B}
  \sum_{\{t_i\}} \delta \left[ t- {\textstyle \sum_{i=1}^4} t_i \right] 
  \sum_{\alpha,\beta} \sum_{s_i,f_i} 
  \Sigma_{\rm TR3a} \times P_{\rm TR3a}
  \times \Delta_{\rm TR3a},
\end{equation}
where 
\begin{eqnarray}
  \label{eq:components_TR3a}
  \Sigma_{\rm TR3a} 
  &=& 
\sigma^{(\alpha)}_{s_4f_3}
\sigma^{(\beta)}_{s_3f_2}
\sigma^{(\alpha)}_{s_2f_1}
\sigma^{(\beta)}_{s_1f_4}
\,
\sigma^{(\alpha)*}_{f_3s_2}
\sigma^{(\beta)*}_{f_2f_4}
\sigma^{(\alpha)*}_{s_4f_1}
\sigma^{(\beta)*}_{s_1s_3}
\,,\\
  P_{\rm TR3a}
  &=& P_{(\beta,s_1) \ra (f_1,\alpha)}^{(t_1-1)}
  P_{(\alpha,s_2) \ra (f_2,\beta)}^{(t_2-1)}
  P_{(\beta,s_3) \ra (f_3,\alpha)}^{(t_3-1)}
  P_{(\alpha,s_4) \ra (f_4,\beta)}^{(t_4-1)}\,.
\end{eqnarray}
If arc 1 is long enough to be ergodic, the sum over $s_1$ simplifies to
\begin{eqnarray}
\label{sum-to-zero}
(1-\de_{s_3 f_4}) \sum_{s_1} \sigma^{(\beta)}_{s_1f_4}
\sigma^{(\beta)*}_{s_1s_3}
=(1-\de_{s_3 f_4})\de_{s_3 f_4}=0\,.
\end{eqnarray}
If arc 2 or arc 3 are ergodic we can carry out a similar sum over $f_2$ or
$f_3$ respectively, these sums also yield the answer zero.  Thus we can only
get a non-zero contribution when arc 4 is the only ergodic path.  However, we
argue in Section~\ref{sect:estimates} that we need at least two arcs to be
long (ergodic) since otherwise the contribution can be neglected.  Thus the
non-zero contribution discussed above will only give a small correction which
will vanish in the limit $B \to \infty$.  The two restrictions $1 \ne \bar{2}$
and $1 \ne 3$ do not change the above argument at all, so we have ignored
them.  We conclude that
\begin{eqnarray}
\label{KTR3a-result}
K_{\rm TR3a}(\tau) = 0\,.
\end{eqnarray}


\subsection{Summation of TR3b}

Here we calculate the contributions of orbits with the topology of
TR3b which obey the conditions $s_2 \ne f_4$, $s_4 \ne f_2$, $1\ne
\bar{1}$ and $3 \ne \bar{3}$.  The first two conditions will be
enforced by means of a factor
\begin{eqnarray}
\Delta_{\rm TR3b} = (1-\de_{s_2 f_4})(1-\de_{s_4 f_2}),  
\end{eqnarray}
the latter two we will enforce below ``by hand''.  Thus
\begin{equation}
  \label{eq:fin_TR3b}
  K_{\rm TR3b}(\tau) = \frac12 \frac{t^2}{B}
  \sum_{\{t_i\}} \delta \left[ t- {\textstyle \sum_{i=1}^4} t_i \right] 
  \sum_{\alpha,\beta} \sum_{s_i,f_i} 
  \Sigma_{\rm TR3b} \times P_{\rm TR3b} \times \Delta_{\rm TR3b} \ ,
\end{equation}
where
\begin{eqnarray}
  \label{eq:components_TR3b}
  \Sigma_{\rm TR3b} 
  &=& 
\sigma^{(\beta)}_{s_4f_3}
\sigma^{(\beta)}_{s_3f_2}
\sigma^{(\alpha)}_{s_2f_1}
\sigma^{(\alpha)}_{s_1f_4}
\,
\sigma^{(\beta)*}_{f_2f_3}
\sigma^{(\beta)*}_{s_3s_4}
\sigma^{(\alpha)*}_{f_4f_1}
\sigma^{(\alpha)*}_{s_1s_2}
\,\\
  P_{\rm TR3b}
  &=& P_{(\alpha,s_1) \ra (f_1,\alpha)}^{(t_1-1)}
  P_{(\alpha,s_2) \ra (f_2,\beta)}^{(t_2-1)}
  P_{(\beta,s_3) \ra (f_3,\beta)}^{(t_3-1)}
  P_{(\beta,s_4) \ra (f_4,\alpha)}^{(t_4-1)}\,.
\end{eqnarray}
We only need to consider cases where $t_1 \ge 3$ and $t_3 \ge 3$ 
because shorter arcs are purely self-retracing ($1 = \bar{1}$) and so 
must be excluded.  We will treat the restrictions $1 \neq \bar{1}$ and
$3 \neq \bar{3}$ using the following inclusion-exclusion procedure: 
the sum in (\ref{eq:fin_TR3b}) with these restrictions is equal to 
the sum without the restrictions, 
minus the sum with $1 = \bar{1}$, 
minus the sum with $3 = \bar{3}$, 
plus the sum with both $1 = \bar{1}$ and $3 = \bar{3}$.

The first sum yields zero after the summation over $s_1$ or over $s_3$ in a
fashion similar to Eq.~(\ref{sum-to-zero}).  The second sum we perform
with respect to $s_3$ while the third is summed with respect to
$s_1$, in both cases the answer is zero.  
Thus $K_{\rm TR3b}(\tau)$ is equal to the sum with 
both $1 = \bar{1}$ and $3 = \bar{3}$, which can be written as
\begin{eqnarray}
  \fl  K_{\rm TR3b}(\tau) = \frac12 \frac{t^2}{B^3}
  \sum_{\{t_i\}} \delta \left[t-{\textstyle \sum_{i=1}^4} t_i\right]
  \nonumber\\
  \times \sum_{\alpha,\beta} \sum_{s_i,f_i} 
  \widetilde{P}_{(\alpha,s_1)\to(s_1,\alpha)}^{(t_1-1)}
  \widetilde{P}_{(\beta,s_3)\to(s_3,\beta)}^{(t_3-1)}
  \times\Delta_{\rm TR3b} \times\Sigma_{\rm TR3b}\times
\delta_{s_1,f_1}\delta_{s_3,f_3},
\end{eqnarray}
where we used the fact that $P_{(\alpha,s_j) \ra (f_j,\beta)}^{(t_j-1)} =
B^{-1}$ for $j = 2, 4$, while  
$\tilde{P}_{(a,b) \ra (b,a)}^{(t)}$ is defined above 
eq.~(\ref{eq:self-retracing}).  
Upon substitutions $f_1=s_1$ and $f_3=s_3$, and using
the symmetry of the matrices $\sigma$, Eq.~(\ref{symsig}), $\Sigma_{\rm TR3b}$
becomes
\begin{equation}
  \Sigma_{\rm TR3b} = 
  |\sigma^{(\alpha)}_{s_1f_4}|^2 |\sigma^{(\alpha)}_{s_2s_1}|^2
  |\sigma^{(\beta)}_{s_3f_2}|^2 |\sigma^{(\beta)}_{s_4s_3}|^2
  = N^{-4}.
\end{equation}
We also notice that the probabilities $\widetilde{P}$ do not depend on the
start and end bonds.  Now we can perform the sum over $\alpha$, $\beta$ and
all $s_i$ and $f_i$, which, taking into account various delta-functions, gives
the factor $N^6(N-1)^2$.  We get
\begin{equation}
  \fl  K_{\rm TR3b}(\tau) = \frac12 \frac{t^2N^2(N-1)^2}{B^3}
  \sum_{\{t_i\}} \delta \left[ t- {\textstyle \sum_{i=1}^4} t_i \right] 
  \widetilde{P}^{(t_1-1)}
  \widetilde{P}^{(t_3-1)},
\end{equation}
from which it is clear that the dominant contribution comes from
$t_1=3,4$ and $t_3=3,4$; the contributions from other values of $t_1$
and $t_3$ are of order $O(N^{-1})$.  After carrying out
the sum over $t_2$ using the $\delta$-function which forces
$t_4=t-t_2-n$ with $n=t_1+t_3= 6,7,8$ we get
\begin{eqnarray}
\label{KTR3b-result}
K_{\rm TR3b} = 4 \times \frac12 {t^2 \over B^3}
\sum_{t_2=3}^{t-3-n} 1 = 2 {t^3 \over B^3} = 2 \tau^3 
\end{eqnarray}
where we have dropped corrections which vanish in the limit $B,N \to
\infty$ and the factor 4 comes from the number of possible choices of $t_1$
and $t_3$.


\subsection{Summation of TR3c}

Here we calculate the contributions of orbits with the topology of TR3c
which obey the following restrictions.  First we should only count {\it
  half} the contribution when $s_2=f_3$ or $s_1=f_2$.  Secondly $s_2 \ne s_3$,
$1\ne \bar{1}$, $2\ne \bar{2}$ and $3\ne \bar{3}$.  The restrictions which
apply to whole arcs will again be enforced ``by hand'' using an
inclusion-exclusion procedure similar to that used above, the rest of
the restrictions are 
\begin{eqnarray}
\Delta_{\rm TR3c} = \left(1-\delta_{s_2s_3}\right) 
\left(1-\half  \delta_{s_2f_3}\right) 
\left(1-\half \delta_{s_1f_2}\right).
\end{eqnarray}
Thus 
\begin{equation}
  \label{eq:fin_TR2}
  K_{\rm TR3c}(\tau) =  \frac{t^2}{B}
  \sum_{\{t_i\}} \delta \left[ t- {\textstyle \sum_{i=1}^3} t_i \right] 
  \sum_{\alpha} \sum_{s_i,f_i} 
  \Sigma_{\rm TR3c} \times P_{\rm TR3c} \times \Delta_{\rm TR3c},
\end{equation}
where 
\begin{eqnarray}
  \label{eq:components_TR2}
  \Sigma_{\rm TR3c} 
  &=& 
\sigma^{(\alpha)}_{s_3f_2}
\sigma^{(\alpha)}_{s_2f_1}
\sigma^{(\alpha)}_{s_1f_3}
\,
\sigma^{(\alpha)*}_{f_2f_3}
\sigma^{(\alpha)*}_{s_3f_1}
\sigma^{(\alpha)*}_{s_1s_2}
\,,\\
  P_{\rm TR3c} 
  &=& P_{(\alpha,s_1) \ra (f_1,\alpha)}^{(t_1-1)}
  P_{(\alpha,s_2) \ra (f_2,\alpha)}^{(t_2-1)}
  P_{(\alpha,s_3) \ra (f_3,\alpha)}^{(t_3-1)}\,.
\end{eqnarray}
The summation here is similar to the sums in TR3b: first we
ignore the restriction $1 \ne \bar{1}$ (but enforce the restrictions 
$2 \ne \bar{2}$ and $3 \ne \bar{3}$) and carry out the sum over
$f_1$ to get
\begin{equation}
  (1-\delta_{s_2s_3}) 
  \sum_{f_1} \sigma^{(\alpha)}_{s_2f_1} \sigma^{(\alpha)*}_{s_3f_1}
  = (1-\delta_{s_2s_3}) \delta_{s_2s_3} = 0\,.
\end{equation}
Then we subtract the sum over orbits with $1 = \bar{1}$ 
(again enforcing the restrictions $2 \ne \bar{2}$ and $3 \ne \bar{3}$).  
Similarly to
TR3b, it turns out that the dominant contribution comes from orbits
with $t_1=3,4$, i.e. $\widetilde{P}^{(t_1-1)} = N^{-2}$.  Since $t_1 \ll
t$ we use the argument from Section~\ref{sect:estimates} to note that we
are only interested in orbits where $t_2,t_3 \sim t$ and thus both arc 2
and 3 are ergodic. This leaves us with
\begin{equation}
  K_{\rm TR3c}(\tau) =  - 2\times \frac{t^2}{B^3N^2}
  \sum_{t_2+t_3 = t-n} 
  \sum_{\alpha} \sum_{s_i,f_i} 
  \Sigma_{\rm TR3c} \times\Delta_{\rm TR3c}\times \delta_{s_1,f_1},
\end{equation}
where $n=3,4$.  We sum over $f_1$ using Eq.~(\ref{symsig}) to get
\begin{equation}
  \Sigma_{\rm TR3c} = \left| \sigma^{(\alpha)}_{s_2s_1} \right|^2
  \sigma^{(\alpha)}_{s_1f_3}  \sigma^{(\alpha)}_{s_3f_2} 
  \sigma^{(\alpha)*}_{f_2f_3} \sigma^{(\alpha)*}_{s_3s_1}
  = N^{-1} \sigma^{(\alpha)}_{s_1f_3}  \sigma^{(\alpha)}_{s_3f_2} 
  \sigma^{(\alpha)*}_{f_2f_3} \sigma^{(\alpha)*}_{s_3s_1},
\end{equation}
open up the brackets in $\Delta_{\rm TR3c}$,
\begin{equation}
  \left( 1-\half \delta_{s_2f_3} \right) 
  \left( 1-\half \delta_{s_1f_2} \right) 
  = \left( 1-\half \delta_{s_1f_2} \right) - \half \delta_{s_2f_3} 
  + {\textstyle{1 \over 4}} \delta_{s_2f_3} \delta_{s_1f_2},
\end{equation}
and are now facing the sum
\begin{equation}
\fl  \sum_{\alpha, s_1,s_2,f_2,s_3,f_3} 
  \sigma^{(\alpha)}_{s_1f_3}  \sigma^{(\alpha)}_{s_3f_2} 
  \sigma^{(\alpha)*}_{f_2f_3} \sigma^{(\alpha)*}_{s_3s_1}
  \left(1-\delta_{s_2s_3}\right) 
  \left[  \left( 1-\half \delta_{s_1f_2} \right) - \half \delta_{s_2f_3} 
  + {\textstyle{1 \over 4}} \delta_{s_2f_3} \delta_{s_1f_2}   \right].
\end{equation}
Invoking the unitarity of the $\sigma$-matrices, it is an easy exercise to
show that this sum evaluates to $N^2(N-1)/2 + N(N-1)/4$.
%

Combining the above information and ignoring subdominant contributions
we arrive at
\begin{eqnarray}
\label{KTR3c-result}
  K_{\rm TR3c}(\tau)
  =  - 2 \frac{t^2}{B^3N^3} \sum_{t_2} N^2(N-1)/2 = -\tau^3.
\end{eqnarray}


\subsection{The TR3 result}

Remembering that we proved (NTR3a + NTR3b)=0 in Section~\ref{sect:sum_NTR3},
we simply need to substitute the results of the three previous subsections
into Eq.(\ref{eq:TR3sum}) to get
\begin{eqnarray}
K_{\rm TR3}(\tau)= 2 \tau^3  \ .
\end{eqnarray}
Combining this result with the one in \cite{BSW02} {\em proves} that the form
factor for the fully-connected Fourier graph coincides with the GOE form
factor up to the third order in $\tau$.


\section{Estimating the order of a diagram}
\label{sect:estimates}

In this section we discuss a rule for finding all diagrams which contribute to
the $n$th order in the small $\tau$ expansion of the form factor. The rule is
\begin{eqnarray}
  \label{eq:rule}
  (\# \hbox{of arcs}) - (\# \hbox{of intersections}) = (n-1).
\end{eqnarray}
Thus for $n=2$, we need only one diagram which is $(2,1)$ in the
format ($\#$of arcs, $\#$of intersections), and this is the
contribution we considered in \cite{BSW02}.  Here we are interested in
$n=3$, so we must consider both $(3,1)$ and $(4,2)$.  It is these
diagrams that we show in Fig.~\ref{fig1}.

To get the rule (\ref{eq:rule}) we count powers of $B$ in a diagram's
contribution.  Equations (\ref{eq:ff-PO-NTR}) and (\ref{eq:ff-PO-TR}) have a
prefactor of $B^{-1}$ so a $\tau^n$-contribution to the form factor must get
$B^{-(n-1)}$ from the summation over the orbits.  In the ergodic limit,
according to Eq.~(\ref{eq:ergodic-tau}), each
arc will contribute the weight $B^{-1}$, while each intersection contributes
the weight $B$, thus we have Eq.~(\ref{eq:rule}).  The origin of the factor of
$B$ associated with each intersection can be explained as follows.  First of
all, the set of all vertices $\{v_j\}$ adjacent to an intersection point
$\gamma$ can be split into two equal subsets satisfying the following
property: if there is a transition $(v_j, \gamma)\ra(\gamma, v_m)$ in either
$P$ or $Q$ then $v_j$ and $v_m$ belong to different 
subsets\footnote[2]{
  In other
  words the graph built on vertices $v_j$, connected if there is a transition
  $(v_j, \gamma)\ra(\gamma, v_m)$, is bipartite.  This graph is nothing else
  but the structure drawn inside the circles in Fig.~\ref{fig1}.  The
  graph is bipartite since it is 2-regular (the valency of each vertex is 2)
  and each connected component contains an even number of bonds.}.  
This is
particularly simple for the NTR3 diagrams where the two sets are simply
$\{s_i\}$ and $\{f_i\}$.
If we now do the summation over all vertices in one subset and invoke the
unitarity of the scattering matrix at the vertex $\gamma$, the result will be
a product of $\de$-functions
$\delta_{u_1u_2}\delta_{u_2u_3}\cdots\delta_{u_ku_1}$ where $u_k$ are the
vertices from the second subset, ordered in an appropriate way.  Now the
summation over $u_2, \ldots, u_k$ will give 1 while the summation over $u_1$
and $\gamma$ will give the sought-after factor of $B$, since the only
restriction on $u_1$ and $\gamma$ is that they have to be two ends of the same
bond.


To make this recipe work for the diagonal term ($\sim\tau^1$), the
corresponding diagram being just a looping arc, we need one extra
ingredient, the starting vertex for the loop.  The position of the
vertex is not determined, it can be placed anywhere on the looping
arc, unlike the intersection points in other diagrams.  To compensate
for this ambiguity when we sum over all periodic orbits fitting such
diagram, we divide the sum by the number of vertices in the loop, $t$.

Now we discuss why counting of powers of $t$ does {\em not} work for obtaining
Eq.~(\ref{eq:rule}).  Let us estimate the order of $t$ for a given diagram.
Firstly, there is $t^2$ in the prefactor of Eq.~(\ref{eq:ff-PO-NTR}) or
(\ref{eq:ff-PO-TR}).  Secondly, for a diagram with $a$ arcs, the lengths $t_i$
of arcs satisfy $\sum_{i} t_i = t$ thus the sum over all possible $t_i$ gives
a factor proportional to $t^{a-1}$.  Then of the diagrams in Fig.~\ref{fig1},
NTR3b and TR3c appear to have four powers of $t$ while the rest have five.
Similarly, the diagram we evaluated in \cite{BSW02} gets three powers of $t$.
The leading contributions to all the diagrams appear to have at least one more
power of $t$ than they should\footnote{This, however, is not the case for the
  ``diagonal'' diagram where the power of $t$ is right, which explains why
  advancing beyond the diagonal approximation was (and still is) so hard.}.
However, we show in \cite{BSW02} and the present article that the numerical
coefficient of this ``out of order'' term is zero---at least for diagrams
contributing up to third order in $\tau$.

The arguments given above in favour of Eq.~(\ref{eq:rule}) are
certainly too vague to be considered a proof.  In particular, we cannot
presently check our assumption that terms giving incorrectly large
powers in $t$ disappear also for more complicated diagrams.  However,
summing rigorously the contributions of all the diagrams obtained from
this set of rules we show {\it a posteriori}\/ that we indeed get an
expansion which depends only on the scaled time $\tau$.  We also take
confidence in our method from the fact that our rule generates the
same diagrams that were used in perturbative calculations of the form
factor for disordered systems with the non-linear sigma model
\cite{SLA98}.

Nevertheless, counting powers of $t$ is very useful in the following
situation: If for some reason the lengths of some arcs are forced to be fixed,
the estimated power of $t$ can drop low enough that we can safely ignore the
contribution of such a diagram in the $B \to \infty$ limit without actually
evaluating it.  In particular, we see that to get a non-vanishing contribution
of order $\tau^3$, at least two arcs in any diagram must have unrestricted
lengths.
Note that this does {\em not} mean that there is no contribution from orbit
pairs where the maximum length of all arcs is restricted. 
For any given $B$ and $t$ there are orbit pairs with so many self-intersections
that the maximum arc length is less than the time required for ergodicity.
Then the method discussed in this paper must fail, this may explain
why the power series expansion in $\tau$ breaks down at
$\tau=1$ for NTR (and at $\tau=1/2$ for TR) despite the fact that the PO sums
Eqs.~(\ref{eq:ff-PO-NTR}), (\ref{eq:ff-PO-TR}) are exact. 


\section{Conclusions}

At first sight, the achievements of the present paper may appear
moderate.  What is the point in going from second to third order in a
series which is infinite, in particular if this step becomes possible
only by restricting the range of systems considered to very special
models? But we believe that such a point of view is too short-sighted.
Semiclassical theories are obviously indispensable for a complete
understanding of spectral statistics in systems with chaotic classical
dynamics. As they are not based on a random matrix conjecture, such
theories have the potential to account for important system-specific
corrections which one can hope to extract once the emergence of
universality within semiclassics is fully understood. On the other
hand, the restriction to the diagonal approximation has so far
severely limited the success of the semiclassical approach.

Going beyond the diagonal approximation in semiclassical PO theories
is possible, as demonstrated by Sieber and Richter \cite{SR01,S02}.
But when more than a first-order correction is required, one will
inevitably encounter the problems discussed in this paper.  For
example, one needs methods to select diagrams contributing at a given
order, and we suggested a solution in Section~\ref{sect:estimates}. It
will be necessary to account for the ambiguity introduced by the
representation of the form factor in terms of diagrams, and we have
solved this problem at least in quantum graphs for the diagrams which
contribute up to third order (Section~\ref{sect:doublecountII}). It is
important to note that a variety of orbits give the relevant
contributions, and our calculations in Section~\ref{sect:sum_TR3}
indicate that the generalization of the leading-order correction to
higher orders cannot be achieved by considering a single type of
diagram only.

Here the third-order result for TR systems is limited to a class of
uniformly hyperbolic quantum graphs. However, we have no reason to
believe that any conclusion will be substantially changed when the
calculation is done for a more generic model as, e.~g.,\ in
\cite{Bpre}.  While going further than third order for TR-systems is
beyond us at the moment, the prospects for doing this in NTR-systems
are more promising.  We hope that the method presented in
Section~\ref{ntr:gen} will prove applicable there.

\section{Acknowledgments}
We gratefully acknowledge extremely productive discussions held with M.~Sieber
and U.~Smilansky.  We also would like to thank the group of Prof F.~Haake in
Universit\"at Essen for hospitality and exchange of ideas.  We are
grateful to a referee who found a large number number of misprints and
inconsistencies in the paper and helped us improve the presentation
of Section~\ref{sect:doublecountI}.  A significant part
of the research was performed while both GB and RW were working at the
Weizmann Institute of Science, Rehovot, Israel. During that time GB was
supported by the Israel Science Foundation, a Minerva grant, and the Minerva
Center for Nonlinear Physics; and RW was supported by the U.S.-Israel
Binational Science Foundation (BSF) and the German-Israel Foundation (GIF).
HS thanks the Weizmann Institute for kind hospitality during a number of
visits.  RW is now supported by an EPSRC Grant.



\end{document}